\documentclass[prl,aps,twocolumn,superscriptaddress,floatfix,showpacs,longbibliography]{revtex4-1}

\usepackage{graphicx}
\usepackage{dcolumn}
\usepackage{bm}
\usepackage{upgreek}

\usepackage[normalem]{ulem}
\usepackage{amsmath,amssymb}
\usepackage[english]{babel}
\usepackage{color}

\makeatletter
\renewcommand{\fnum@figure}{Fig. \thefigure}
\makeatother

\begin{document}

\title{Gate tuning from exciton superfluid to quantum anomalous Hall in van der Waals heterobilayer}

\author{Qizhong Zhu}
\affiliation{Department of Physics and Center of Theoretical and Computational Physics, University of
Hong Kong, Hong Kong, China}

\author{Matisse Wei-Yuan Tu}
\affiliation{Department of Physics and Center of Theoretical and Computational Physics, University of
Hong Kong, Hong Kong, China}

\author{Qingjun Tong}
\affiliation{Department of Physics and Center of Theoretical and Computational Physics, University of
Hong Kong, Hong Kong, China}

\author{Wang Yao}
\email[]{wangyao@hku.hk}
\affiliation{Department of Physics and Center of Theoretical and Computational Physics, University of
Hong Kong, Hong Kong, China}

\date{\today}

\begin{abstract}
Van der Waals heterostructures of 2D materials provide a powerful approach towards engineering various quantum phases of matters.
Examples include topological matters such as quantum spin Hall (QSH) insulator, and correlated matters such as exciton superfluid.
It can be of great interest to realize these vastly different quantum matters on a common platform, however, their distinct origins tend to restrict them to material systems of incompatible characters. Here we show that heterobilayers of two-dimensional valley semiconductors can be tuned through interlayer bias between an exciton superfluid (ES), a quantum anomalous Hall (QAH) insulator, and a QSH insulator. The tunability between these distinct phases results from the competition of Coulomb interaction with the interlayer quantum tunnelling that has a chiral form in valley semiconductors.  
Our findings point to exciting opportunities for harnessing both protected topological edge channels and bulk superfluidity in an electrically configurable platform. 
\end{abstract}

{\maketitle}

\textbf{Introduction}

In exciton Bose-Einstein condensate, an electron and a hole pair into an exciton that can flow without dissipation. Confining electron and hole into two separate layers allows the exciton superfluid to manifest as counter-flowing electrical supercurrents in the electron and hole layers \cite{eisenstein_2004,nandi_exciton_2012}. 
Van der Waals (vdW) heterostructures are ideal realisations of the double-layer geometry for exploring this correlated phase of matter driven by Coulomb interaction \cite{liu_quantum_2017,li_excitonic_2017,burg_strongly_2018}. Evidences of the counterflow supercurrents in the quantum Hall regime are recently reported in graphene double-bilayers \cite{liu_quantum_2017,li_excitonic_2017}. High-temperature exciton superfluid phases in absence of magnetic field are also predicted in graphene \cite{min_room-temperature_2008} and transition metal dichalcogenides (TMDs) double-layer heterostructures \cite{wu_theory_2015,fogler_high-temperature_2014}. 

Quantum spin Hall (QSH) insulators are topological state of matter driven by the spin-orbit coupling, a single-particle relativistic effect \cite{qi_quantum_2009,hasan_textitcolloquium_2010}. In 2D crystals and their vdW heterostructures, the miniaturisation in thickness can lead to remarkable gate-tunable QSH phase, featuring helical edge states that can be electrically switched on/off inside the bulk gap \cite{qian_quantum_2014,liu_switching_2015,tong_topological_2017,wu_observation_2018}. 
Electron flow in the helical QSH edge channel is protected from backscattering, except by the spin-flip scatters. Coupling QSH insulator to local magnetic moment in ferromagnetism can suppress the topological order in one spin species \cite{weng_quantum_2015,chang_experimental_2013}, turning QSH into the quantum anomalous Hall (QAH) insulator. QAH features chiral edge state that is completely lossless with the absence of backward channel.
The edge conducting channels of these topological matters, as well as the bulk supercurrents in the exciton superfluid, can have profound consequences in quantum electronics~\cite{eisenstein_2004,nandi_exciton_2012,liu_quantum_2017,li_excitonic_2017,fei_edge_2017,
tang_quantum_2017}.

Here we show the possibility of realizing these vastly different quantum matters with gate switchability on a single platform of TMDs heterobilayer.
What makes this system unique is the coexistence of strong Coulomb interaction that favors spontaneous $s$-wave interlayer electron-hole coherence, and a chiral interlayer tunnelling that creates/annihilates electron-hole pair in the $p$-wave channel only. Their competition leads to a rich phase diagram when the heterobilayer band alignment is tuned towards the inverted regime through the interlayer potential difference induced by the gate (i.e. interlayer bias). At relatively strong dielectric screening, the bias drives transitions from a normal insulator to three nontrivial phases sequentially: (i) exciton superfluid (ES); (ii) coexistence of QAH in spin-up and ES in spin-down species (QAH-ES); and (iii) QSH insulator. At weak screening, magnetic order spontaneously develops along with the interlayer coherence, where the heterobilayer can be gate tuned between (iv) a magnetic ES (MES), and (v) a QAH phase. Remarkably, the topologically distinct phases are connected without gap closing, but through spontaneous symmetry breaking instead. The gate switchability, together with the sizable QSH/QAH gap that can exceed room temperature, point to practical spintronic highways at the electrically reconfigurable topological interfaces.

\begin{figure*}
\includegraphics[width=15cm]{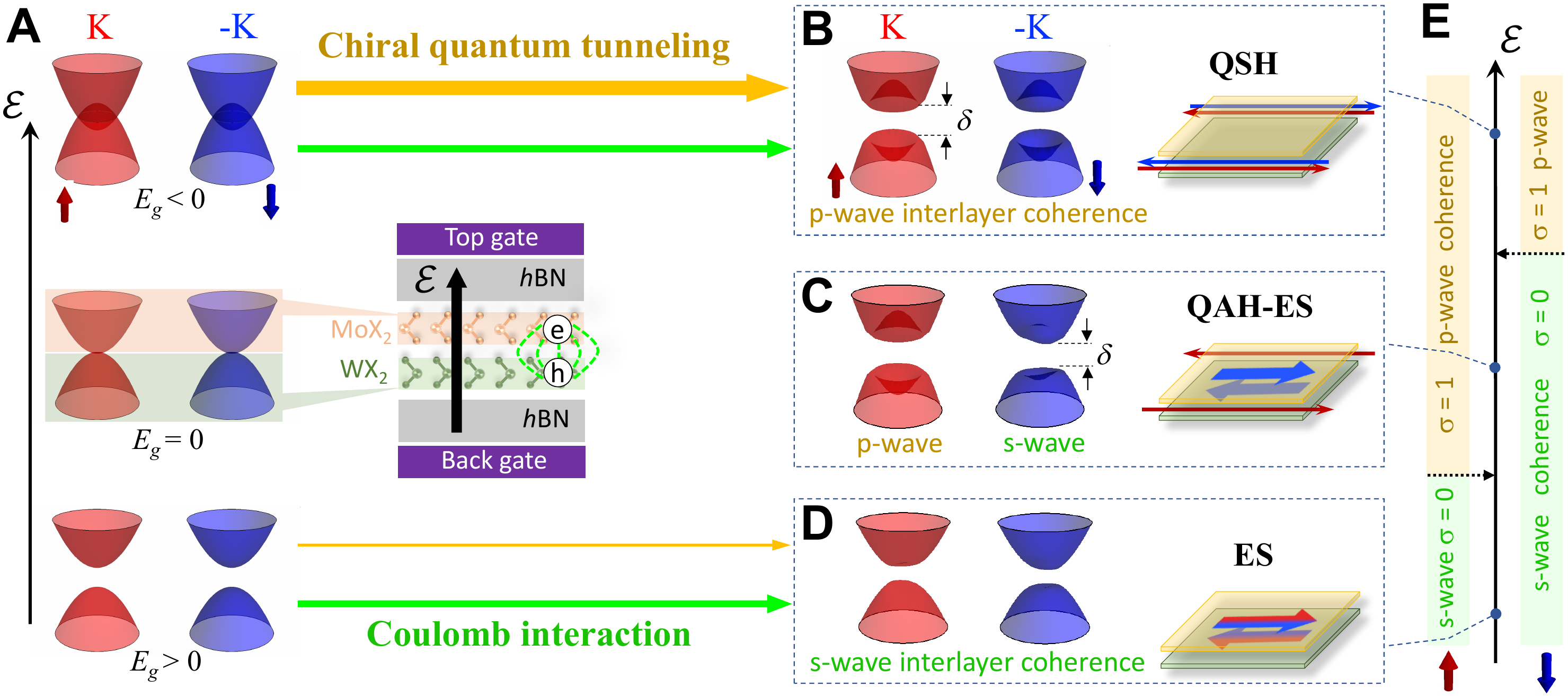}\newline
\caption{\textbf{Band inversion under competition between Coulomb interaction and chiral quantum tunnelling}. (\textbf{A}) Heterobilayers of semiconducting TMDs feature the type-II band alignment, where conduction and valence band edges are spin-valley locked massive Dirac cones from opposite layers. The band gap $E_g$ in the non-interacting limit can be closed and inverted by an interlayer bias.  (\textbf{B}-\textbf{D}) Phases of the bilayer under the competition between Coulomb and interlayer quantum tunnelling of the symmetry dictated chiral form (c.f. text). 
The dominance of Coulomb at positive small $E_g$ leads to exciton superfluid (ES) of spontaneous $s$-wave interlayer coherence, shown in \textbf{D}. The dominance of chiral quantum tunnelling at negative $E_g$ pins the interlayer coherence in the $p$-wave channel, where the bilayer is a QSH insulator, shown in \textbf{B}. (\textbf{E}) Such phase transition occurs non-simultaneously for spin up and down species. Between ES and QSH phases, there is a phase of coexistence of ES in spin-down and QAH in spin-up species, as shown in \textbf{C}.}
\label{figure1}
\end{figure*}

\textbf{Results}

Figure 1 schematically explains the gate-controlled phase transitions. TMD heterobilayers have the type-II band alignment where the conduction (valence) band edge consists of upper (lower) massive Dirac cones from the top (bottom) layer, at the K and -K corners of hexagonal Brillouin zone. Because of the spin-valley locking in TMDs monolayers, only the spin up (down) massive Dirac cones are relevant at the K (-K) valley~\cite{tong_topological_2017}. At small or negative $E_g$ (bandgap), a pair of layer-separated electron and hole can be spontaneously generated by their Coulomb interaction, or by the interlayer quantum tunnelling. For several high symmetry stacking configurations, the $C_3$ rotational symmetry dictates the tunnelling to have a chiral dependence on the in-plane wavevector ($t \propto k_x \pm  i k_y$)~\cite{tong_topological_2017}. In the inverted regime ($E_g<0$), quantum tunnelling becomes resonant at $k \sim \sqrt{|E_g|}$, so its effective strength grows with $|E_g|$, the latter becoming a knob to control the dominance between Coulomb interaction and quantum tunnelling. Major features of the phase diagram can then be intuitively anticipated.

When an interlayer bias tunes $E_g$ towards the inverted regime, the Coulomb interaction first drives the heterobilayer into exciton superfluid with spontaneous $s$-wave interlayer coherence (Fig. 1D), as well-studied in TMDs double-layer designed with interlayer tunnelling quenched~\cite{wu_theory_2015,fogler_high-temperature_2014}. 
In contrast to conventional double-layers where tunnelling will fix the phase of the interlayer coherence~\cite{eisenstein_2004,nandi_exciton_2012}, here moderate tunnelling of the unique $p$-wave form  does not affect excitons condensed in the $s$-wave channel. Instead, the chiral tunnelling induces a background coherence in the $p$-wave channel, whose interference with the condensate in $s$-wave channel enables {\it in situ} measurement on the condensate phase through an in-plane electrical polarization. In such case, the ES phase becomes nematic, with a spontaneous breaking of the rotational symmetry.  

Only deep in the inverted regime, the eventual dominance of quantum tunnelling pins the interlayer coherence entirely in the $p$-wave channel, and the heterobilayer becomes a QSH insulator. Helical edge states appear in the hybridization gap $\delta$ (Fig. 1B), the magnitude of which is significantly enhanced by Coulomb interaction compared to the non-interacting case~\cite{tong_topological_2017}. 
The phase transition between ES and QSH does not happen simultaneously for spin up and down species (Fig. 1E), leaving a bias range for the coexistence of exciton superfluidity in one spin species, and quantum anomalous Hall in the other. This QAH-ES phase features both counterflow bulk supercurrent and chiral edge state in the bulk gap (Fig. 1C). 

In the non-interacting limit, the effect of chiral quantum tunnelling in TMDs heterobilayer is well described by the two-band Hamiltonian \cite{tong_topological_2017}: $\hat{H}_{0, \tau}=\sum_\mathbf{k}\left( \hat a_{\mathbf{k}}^\dagger, \hat b_{\mathbf{k}}^\dagger \right)\big[\eta k^2+\varepsilon_{\mathbf{k}}\sigma_z+t_{\tau\mathbf{k}}\sigma_+ + t^*_{\tau\mathbf{k}}\sigma_- \big]\left( \hat a_{\mathbf{k}}, \hat b_{\mathbf{k}}\right)^T$, where $\hat a_{\mathbf{k}}^\dagger$ ($\hat b_{\mathbf{k}}$) creates electron (hole) in top (bottom) layer, $\tau=\pm1$ the valley index, \(\sigma\) the Pauli matrices in layer pseudospin space, and $\varepsilon_{\mathbf{k}}=\hbar^2 k^2/2m+E_g/2$. $m$ is twice the reduced mass of electron and hole, and the $\eta k^2$ term accounts for their mass difference.  The interlayer tunnelling $t_{\tau \mathbf{k}}$ has a stacking-dependent form. For the example of 2H-stacking for epitaxially grown heterobilayer~\cite{hsu_negative_2018}, we have $t_{\tau \mathbf{k}}=v (\tau k_x - i k_y) t^*_{vv}/M$ ~\cite{tong_topological_2017}. $t_{vv}$ is the hopping amplitude between the valence band-edges of the two layers, $v$ the Dirac cone Fermi velocity and $M$ the band gap of monolayer TMD. The ground state $|\Psi \rangle=\prod_\mathbf{\tau k}(u_{\tau \mathbf{k}}+v_{\tau \mathbf{k}} a_\mathbf{k}^\dagger b_\mathbf{k})| 0 \rangle$ then features a $p$-wave interlayer coherence: $u^{*}_{\tau \mathbf{k}}v_{\tau \mathbf{k}} \propto -\tau k_x + i k_y$, where the spin Hall conductivity jumps from $0$ to $1$ at $E_g=0$.

\begin{figure*}
\includegraphics[width=13cm]{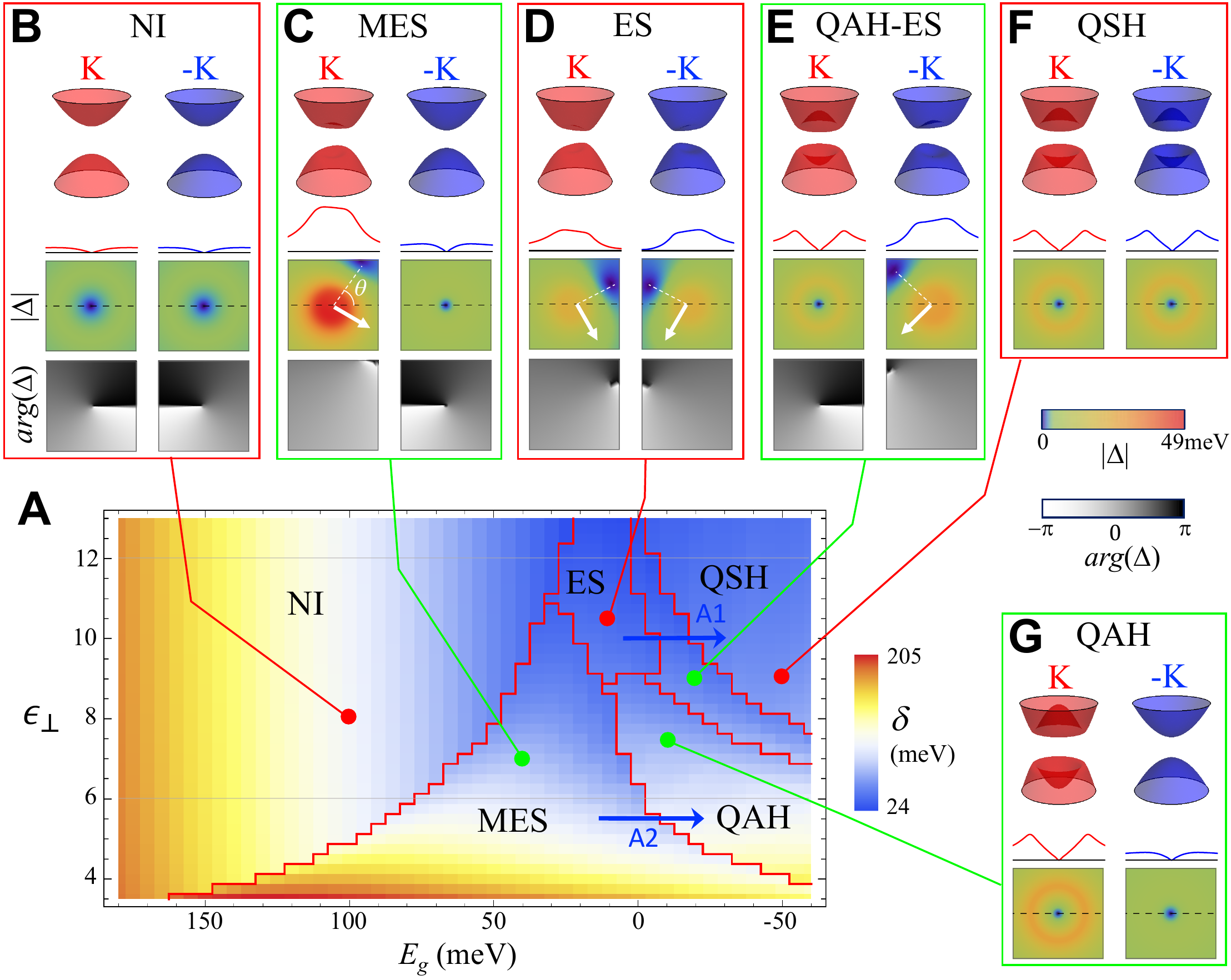}\newline
\caption{\textbf{Phase diagram}. (\textbf{A}) Phase diagram as a function of band gap $E_g$ and interlayer dielectric constant $\epsilon_\perp$. (\textbf{B}-\textbf{G}) Examples of the six phases. The quasiparticle energy bands are shown together with the magnitude $|\Delta|$ and phase angle $\arg(\Delta)$ of the order parameter (see text), over a momentum space region of $[-\frac{\pi}{8a_0}, \frac{\pi}{8a_0}]\times[-\frac{\pi}{8a_0}, \frac{\pi}{8a_0}]$ at the two valleys respectively, with $a_0$ being the lattice constant. The curves atop of $|\Delta|$ map show their value along the dashed cut. The QSH and QAH phases have the same $p$-type $\arg(\Delta)$ map as in the NI phase. In (C), (D) and (E), the exciton density of the ES is $0.019 a_B^{-2}$, $0.028 a_B^{-2}$ and $0.029 a_B^{-2}$  ($a_B \equiv \hbar^2\epsilon/me^2$) respectively, and the anisotropic $|\Delta|$ corresponds to an in-plane electric dipole of $6.8  e\mathrm{\AA} $, $9.0 e\mathrm{\AA} $ and $8.3  e\mathrm{\AA} $ per exciton, in directions denoted by the white arrows. 
}
\label{figure2}
\end{figure*}

Coulomb interaction is well accounted for in double-layer geometry by the Hartree-Fock approximation, as adopted in various studies of quantum phases therein~\cite{zhu_exciton_1995,seradjeh_exciton_2009,pikulin_interplay_2014,budich_time_2014, wu_theory_2015,xue_time-reversal_2018}. 
The electron energy is dressed by the interaction with the electron-hole pairs in the ground state $|\Psi \rangle$. The effective interlayer tunnelling also gets renormalized by the Coulomb interaction, becoming dependent on the electron-hole coherence in $|\Psi \rangle$. The mean-field interacting Hamiltonian reads: $\hat{H}_{\tau}=\sum_\mathbf{k}\left( \hat a_{\mathbf{k}}^\dagger, \hat b_{\mathbf{k}}^\dagger \right)\big[\eta k^2+\xi_{\tau \mathbf{k}}\sigma_z+\left( (-\Delta_{\tau\mathbf{k}} + t_{\tau\mathbf{k}})\sigma_+ + h.c. \right)\big]\left( \hat a_{\mathbf{k}}, \hat b_{\mathbf{k}}\right)^T$,
with $\Delta_{\tau\mathbf{k}}=\sum_{\mathbf{k}'}V_\mathrm{inter}\left(\mathbf{k}-\mathbf{k}'\right)u^{*}_{\tau\mathbf{k}'}v_{\tau\mathbf{k}'}$, and $\xi_{\tau\mathbf{k}} \equiv \varepsilon_{\mathbf{k}}-\sum_{\mathbf{k}'}V_\mathrm{intra}\left(\mathbf{k}-\mathbf{k}'\right)|v_{\tau\mathbf{k}'}|^{2}+e^2 C^{-1} \sum_{\tau' \mathbf{k}'}|v_{\tau' \mathbf{k}'}|^{2}/2 $.  Here $V_\mathrm{intra}$ and $V_\mathrm{inter}$ are the intra- and inter-layer Coulomb interactions respectively. The last term in $\xi_{\tau\mathbf{k}}$ is the classical charging energy of the bilayer as a parallel-plate capacitor, with $C \equiv e^2/[2(V_\mathrm{intra}(0) -  V_\mathrm{inter}(0))]$ being the capacitance per unit area. 
The ground state $|\Psi \rangle$ shall now be solved from the self-consistent gap equation,
\begin{equation}
\Delta_{\tau\mathbf{k}}=\sum_{\mathbf{k}'}V_\mathrm{inter}\left(\mathbf{k}-\mathbf{k}'\right)\frac{\Delta_{\tau\mathbf{k}'} - t_{\tau\mathbf{k}'}}{2\sqrt{\xi_{\tau\mathbf{k}'}^{2}
+|\Delta_{\tau\mathbf{k}'} - t_{\tau\mathbf{k}'}|^{2}}}. \label{gapeq}
\end{equation}
This mean-field approach describes well the exciton condensate in TMDs double-layers with interlayer tunnelling quenched~\cite{wu_theory_2015}.

Figure 2 shows the phase diagram as a function of dielectric constant and band gap $E_g$, calculated from Eq. (1) (see Materials and Methods). Different phases are identified from their distinct interlayer coherence $\Delta_{\tau\mathbf{k}}$ and Hall conductance in the quasiparticle gap. 
At large and positive $E_g$, a small electron-hole coherence $\Delta_{\tau\mathbf{k}}$ is induced in the $p$-wave channel by the quantum tunnelling at large detuning (Fig. 2B), where the bilayer is a normal insulator (NI). When $E_g$ is reduced below
the exciton binding energy, there is a sudden switch-on of the $s$-wave interlayer coherence by the Coulomb interaction. The bilayer is still topologically trivial, but developes the ES either with or without spontaneous magnetic order (Fig.~2C or 2D). Both ES phases have been predicted in TMDs double-layers with quenched tunnelling~\cite{wu_theory_2015}, and the inclusion of chiral quantum tunnelling here introduces little change on the phase boundaries between them and the NI phase. 

We find that the interlayer coherence in the ES ground state is a superposition of $s$-wave and $p$-wave components: $\Delta_{\tau\mathbf{k}}=\tau \Delta_s(\mathbf{k})  e^{-i\tau\theta} -  \tau \Delta_p(\mathbf{k})  e^{-i\tau\phi(\mathbf{k})}$, $\phi(\mathbf{k})$ being the azimuth angle of $\mathbf{k}$, and $\Delta_{s,p}$ are real and positive. The interference leads to a node in $\Delta_{\tau\mathbf{k}}$ at azimuth angle equal to $\theta$ (c.f. Fig.~2C). The phase $\theta$ of the $s$-wave component is unrestricted, so spontaneous symmetry breaking due to the Coulomb interaction still occurs. The order parameters corresponding to different $\theta$ values are related by the operation $\mathcal{G}(\theta) \equiv e^{-i\tau\theta}\times\mathcal{R}(\theta)$, that is, the gauge transformation plus a spacial rotation by angle $\theta$. This is a U(1) symmetry possessed by both Coulomb interaction and chiral tunnelling. 
Consequently, superfluidity is unaffected even when tunnelling is quite significant. Remarkably, such ES features an in-plane electric polarization of azimuth angle $\theta-\tau\pi/2$ (white arrows in Fig.~2C-E), from the interference between the $s$-wave and $p$-wave components of $\Delta_{\tau\mathbf{k}}$. This makes possible the direct observation of the condensate phase $\theta$. 

The spontaneous magnetic order in the exciton condensate arises from a negative exchange interaction between the interlayer excitons~\cite{wu_theory_2015,combescot_effects_2015}. The intralayer and interlayer Coulomb interactions can be grouped into a repulsive dipole-dipole interaction, and an exchange interaction between excitons of same spin/valley only. The exciton exchange interaction is sensitive to the ratio between the interlayer distance and the exciton Bohr radius, and can have a sign change as a function of this ratio \cite{ciuti_role_1998}. The Bohr radius is proportional to the dielectric constant $\epsilon$. At fixed interlayer distance $d$, the exchange interaction can then change from a repulsive one at large $\epsilon$ that favors an unpolarized condensate, to an attractive one at small $\epsilon$ that favors a spin-polarized condensate \cite{wu_theory_2015,combescot_effects_2015}.
The boundary between the spin polarized and unpolarized ES phases is consistent with that found in TMDs double-layer of quenched tunnelling~\cite{wu_theory_2015}. Our calculations show that this boundary can be extrapolated to divide the rest part of the phase diagram at higher excitonic density. At large $\epsilon$ are phases with spin balanced electron-hole density, and at small $\epsilon$ are spin-polarized phases (Fig. 2).

With $E_g$ decreasing into the inverted regime, there is a general trend for the interlayer coherence to switch from the Coulomb favored $s$-wave to the tunnelling favored $p$-wave channel, which is a topological phase transition. In the spin balanced regime, this transition sequentially happens in spin up and down species (c.f. Fig.~1E), changing the bilayer from the ES, to the QAH-ES, and then to the QSH phase (arrow A1 in Fig. 2A). In the spin polarized regime, the topological phase transition in the majority spin species changes the bilayer from the magnetic ES to QAH (arrow A2 in Fig. 2A). The ES and QAH-ES phase regions both shrink with the increase of $\epsilon$, showing the right trend towards a direct transition between NI and QSH phases in the infinite $\epsilon$ (non-interacting) limit. 

\begin{figure}
\includegraphics[width=\linewidth]{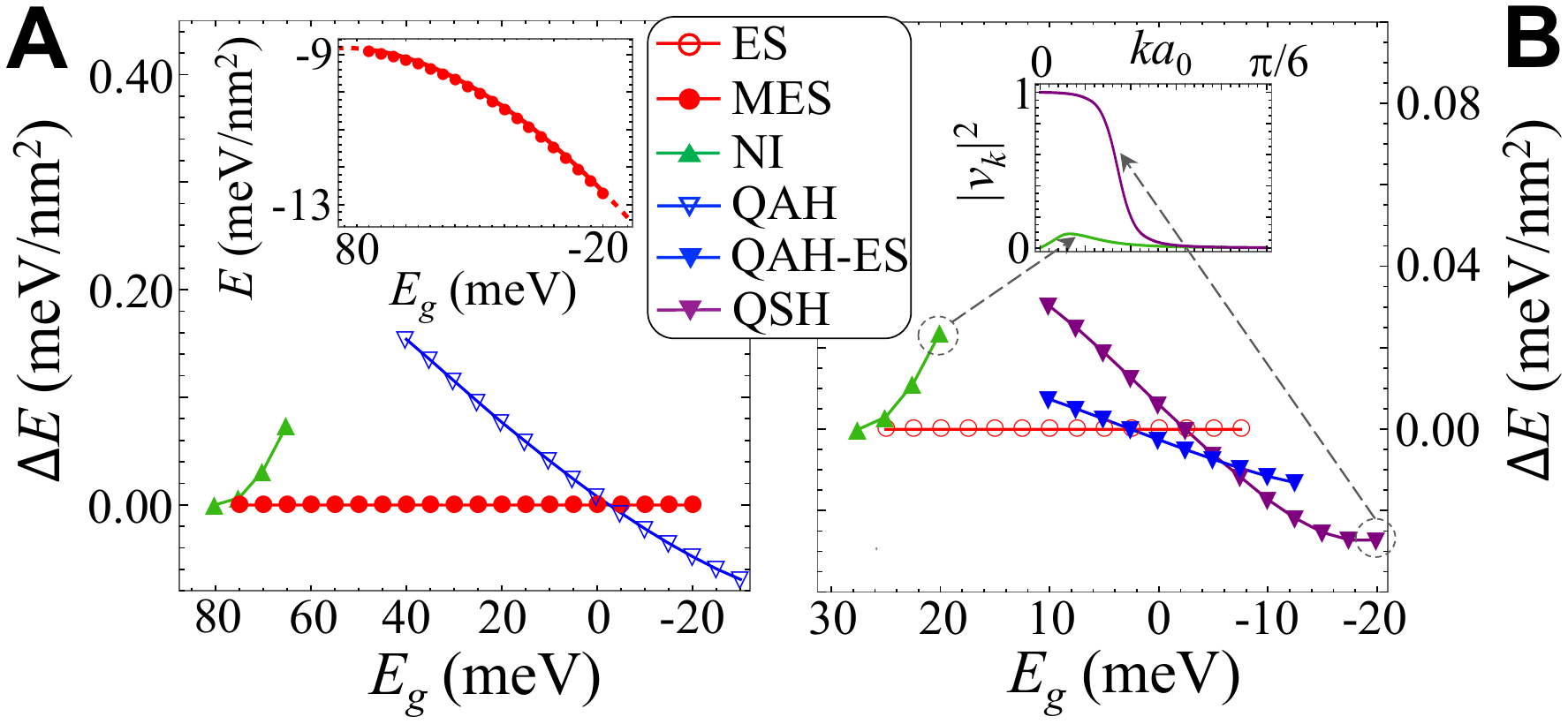}\newline
\caption{\textbf{Topological phase transitions without gap closing}.
(\textbf{A}) Energies of stable solutions of the mean-field Hamiltonian relative to that of the MES state. Inset shows the MES state energy, with dashed part being the extrapolation. $\epsilon_\perp = 6$, corresponding to the lower gray horizontal line in Fig. 2A. (\textbf{B}) Energies of stable solutions measured from the energy of the ES state. $\epsilon_\perp = 12$, corresponding to the upper gray horizontal line in Fig. 2A. The NI and topologically nontrivial QSH (QAH) ground states are  connected, without gap closing, through the ES (MES) ground state with spontaneous symmetry breaking. Inset of \textbf{B} plots the electron-hole pair density $|v_k|^2$ for two representative NI and QSH states. }
\label{figure3}
\end{figure}

\vspace{10 pt}

\textbf{Discussions}

It is important to point out that a sizable quasiparticle gap remains across all phase regions in Fig. 2A, including the boundaries between topologically distinct phases. This is in contrast to the necessary gap closing in topological phase transitions in the non-interacting limit.
Here the NI phase and the topological nontrivial QSH (QAH) phase is connected through the ES (MES) phase with the spontaneous symmetry breaking. The change of topological number in the ground state is accompanied by the symmetry change, and the gap-closing requirement therefore does not apply~\cite{ezawa_topological_2013,yang_topological_2013,pikulin_interplay_2014,xue_time-reversal_2018}. Fig. 3 plots the relative energies of the stable solutions of the mean-field Hamiltonian. Both the discontinuity in the first derivative of energy and multiple stable states close to the transition point show that they are first-order quantum phase transitions. Towards the right end of the ES regions in Fig. 2A, the electron-hole pair density from our calculations is approaching the Mott density~\cite{maezono_excitons_2013}, so likely other correlated phases such as electron-hole plasma can emerge, which is beyond the scope of the mean field approximation here.

The chiral form of the tunnelling, ensured here by the three-fold rotational symmetry of the heterobilayer lattice, is key to the gate tunable phases. When the stacking has some deviation from the high symmetry ones considered, the tunnelling can have an $s$-wave component, which can shrink quantitatively the topological phase regions. Besides, in the presence of $s$-wave tunnelling, as well as the weak trigonal warping effects in TMDs, the interlayer coherence in the ES phase is not completely spontaneous. These two effects can explicitly break the Hamiltonian's U(1) symmetry under $\mathcal{G}(\theta)$. Similar to the role of the interlayer tunnelling in conventional double-layer exciton superfluid~\cite{su_how_2008}, they will lift the ground state degeneracy. Consequently, the Goldstone bosons will not be massless, but remain relatively light if the trigonal warping and the $s$-wave tunnelling component are not large. 

It is also interesting to note that the distinct topological orders of the NI and QSH (QAH) states are reflected in the electron-hole pair density $|v_{\bf k}|^2$, while their $\Delta_{\bf k}$ plots look the same (Fig. 2). As shown in Fig. 3B, $|v_{\bf k}|^2$ plot of the NI state is of the character of BEC type state of tightly bound electron-hole pairs of $p$-orbital relative motion. In contrast, $|v_{\bf k}|^2$ of the QSH state is of the character of BCS state of weak-pairing. This is consistent with earlier work 
showing that the distinction between BEC and BCS in the $p$-wave channel is topological \cite{read_paired_2000}. 

\begin{figure}
\includegraphics[width=\linewidth]{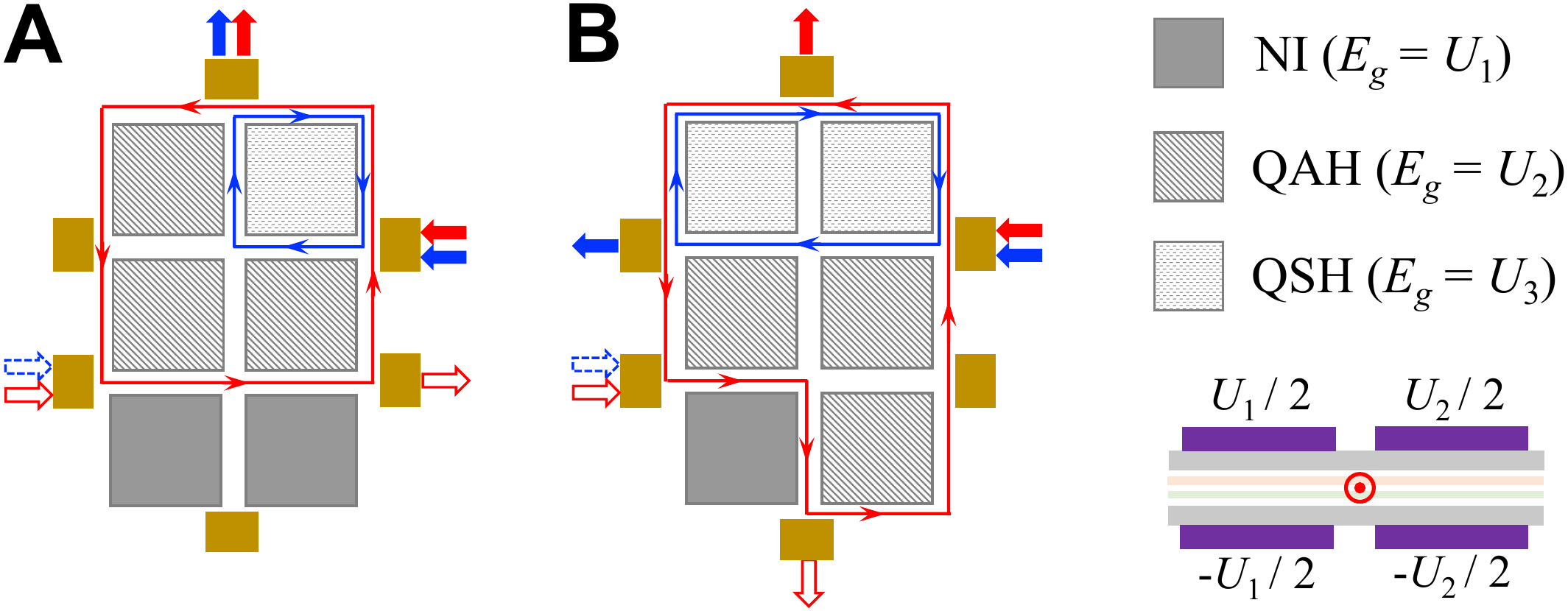}\newline
\caption{\textbf{Configurable spintronics highways}. Topological boundaries between NI, QAH and QSH can be electrically patterned and reconfigured on a bilayer using split top-bottom gate pairs, for wiring protected helical/chiral spin channels. Red and blue color denote spin up and down channels respectively.}
\label{figure4}
\end{figure}
  
The heterobilayers can be formed with a variety of semiconducting TMD compounds that feature similar band structures \cite{tong_topological_2017}, while their different work functions lead to choices on the bandgap. Heterobilayers of MoS$_2$, MoSe$_2$, WS$_2$ and WSe$_2$ have been extensively studied for interlayer excitons in the type-II band alignment \cite{rivera_interlayer_2018}. These heterobilayers have a gap of $1\sim 2$ eV, which requires a large electric field to invert. First principle calculations show that using compounds such as 1H WTe$_2$, CrS$_2$, CrSe$_2$ and CrTe$_2$ as one or both building blocks leads to much smaller gap in the absence of electric field \cite{ras_com_2015,ozcelik_band_2016,tong_topological_2017}, which can be more favorable choices for device applications, allowing heterobilayers to be tuned in the desired regime by a small electric field.

The gate switchability and the sizeable gap that can exceed room temperature in the QSH and QAH phases point to exciting opportunity towards practical  quantum spintronics exploring the protected edge states. 
Using the split top-bottom gate design that has been implemented in bilayer graphene to define valley channels~\cite{martin_2008,li_gate-controlled_2016,qiao_electronic_2011}, topological boundaries between NI, QAH and QSH can be programmed on the heterobilayer for wiring the helical/chiral channels to conduct spin currents, as Fig. 4 illustrates. There also lies an intriguing possibility of integrating these topological channels with the counter-flow superfluidity when the top and bottom layers are separately contacted. 

\textbf{Materials and Methods}

In the numerical calculation of the phase diagram and phase transitions presented in Fig. 2 and 3, we adopt the typical forms of the intra- and inter-layer Coulomb interactions~\cite{wu_theory_2015}: $V_\mathrm{intra}(\mathbf{k})=2\pi e^2/ (\epsilon k) $ and $V_\mathrm{inter}(\mathbf{k})=V_\mathrm{intra}(\mathbf{k})e^{-kd}$. $\epsilon=\sqrt{\epsilon_\parallel\epsilon_\perp}$, where $\epsilon_\parallel$ ($\epsilon_\perp$) is the intralayer (interlayer) dielectric constant. $d=D\sqrt{\epsilon_\parallel/\epsilon_\perp}$, $D$ being the geometric interlayer distance. The chiral tunneling $t_{\tau \mathbf{k}}=v (\tau k_x - i k_y) t^*_{vv}/M$. The parameter values $D= 0.62$ nm, $\epsilon_\parallel/\epsilon_\perp =1.6$, $m=0.5 m_0$ ($m_0$ being electron bare mass), $t_{vv} = 14.4$ meV,  $v=3.512$ eV$\cdot\mathrm{\AA}$, $M=1.66$ eV are used here, based on first principle calculations~\cite{tong_topological_2017,wang_interlayer_2017,kumar_tunable_2012,xiao_coupled_2012}. The valley-coupled gap equation Eq.~(\ref{gapeq}) is numerically solved by convergence to stable solutions with various initial trial \(\Delta_{\tau\mathbf{k}}\).

\textit{Acknowledgment}. We thank Shizhong Zhang and L.-A. Wu for helpful discussions. The work is supported by the Croucher Foundation (Croucher Innovation Award), the RGC (HKU17302617) and UGC (AoE/P-04/08) of HKSAR.

\end{document}